# PSYCHOLOGICAL MODEL OF THE INVESTOR AND MANAGER BEHAVIOR IN RISK

O. A. Malafeyev[1], A. N. Malova[2], A. E. Tsybaeva[3]

St.Petersburg State University, 7/9 Universitetskayanab., St. Petersburg, 199034 Russia

**Abstract**

All people have to make risky decisions in everyday life. And we do not know how true they are. But is it possible to mathematically assess the correctness of our choice? This article discusses the model of decision making under risk on the example of project management. This is a game with two players, one of which is Investor, and the other is the Project Manager. Each player makes a risky decision for himself, based on his past experience. With the help of a mathematical model, the players form a level of confidence, depending on who the player accepts the strategy or does not accept. The project manager assesses the costs and compares them with the level of confidence. An investor evaluates past results. Also visit the case where the strategy of the player accepts the part.

[1] * o.malafeev@spbu.ru
[2] anastasya.malova97@yandex.ru
[3] aleksandra_tsybaeva@mail.ru

## 1. Overview

We will present examples of works showing that confidence in the events depends on experience.

A person's psychological state may affect performance in different ways. Claude Steele and Joshua Aronson (1995) [61] suggest that stress can impair performance. In the final exams, black and white were asked many difficult verbal tasks in two conditions: the first was described as diagnostic, which should be interpreted as an assessment of the individual, while the second was described as a definition of how people solved problems. Steele and Aronson interpret the diagnostic state as putting black subjects at risk of fulfilling a racial stereotype about their intellectual abilities and causing self-doubt and concern about conformity with this stereotype. Negros manifested themselves more weakly when stress was induced (diagnostic condition) than in a neutral state.

Aronson et al. (1999) [62] speaks of a similar effect in white-skinned men. When it was said before the test that there is a stereotype that Asian students are better than Caucasian students in mathematical abilities, these experienced white men passed the test much worse than students who were not told about it. This shows how a person's past stories affect his self-confidence.

Oliver Compte and Andrew Postlevaite (2001) [64] talked about how the psychological state of a person could affect his performance. For example, fear can lead to failure, while memories of past successes gave confidence in what happened.

Simon Baker et al. (1997) [63] also consider positive and negative natures in humans and show that induced mood affects the subject's verbal fluency. In addition, they measure regional cerebral blood flow using positron emission tomography (PET). They find that elevated mood is associated with the activation of areas of the brain associated with different emotions. This conclusion is very interesting because it points to the obvious physiological effects of mood.

## 2. Introduction

A person is subject to different emotions while making a decision, which depends on many factors. One of them is the memory of past events. Bad experience can cause uncertainty in the correctness of the decision that would entail the rejection of this strategy. Also, there may be a reverse situation: success in the past encourages a person to make a decision.

There is evidence that a person is inclined to memorize more negative events of his life, rather than positive ones. For example, the pain of losing a small amount of money will be remembered longer than the joy of winning this amount in the lottery. That is why this article aims to show how life experience influences decision making.

One of the most illustrative examples of this problem is investing in projects. In this situation, a decision is made whether to invest money in a project or not, since there is a large share of the risk of losing the invested amount without having received any profit in the future. Further in the article we will consider the choice of an investor in the conditions of risk and draw up a game-theoretic model.

# 3. Formulation of the problem

Let us analyze how the Investor forms the choice of his strategy. Let A denote the Investor's strategy, which implies a contribution to the project. Adopting strategy A, there is both a risk and an opportunity to make a profit. As a rule, the greater the risk of investing, the greater the expected return.

The question arises, what (logical, psychological) mechanism can lead an agent to the most effective outcome of this problem? This question remains one of the central in modern game theory.

We assume that the main problem of acceptance is confidence, or rather its absence. Certainly, confidence can be influenced by past experience of the Investor: which of his investments justified themselves and which of them did not.

In this model, we believe that it is the past experience that affects confidence, and thus the likelihood of the investor getting the maximum profit.

**Formalization of the game-theoretic model of strategy choice under risk conditions**

**Investor strategy selection. Formation of the level of investor confidence.**

The beginning of the procedure for concluding a potential transaction will correspond to the strategy chosen by the Investor A, which will mean in the future that the Investor has decided to invest in the project. Let us analyze how the Investor forms the choice of his strategy.

Let $p^0 \in [0,1]$ be the Investor's confidence level in the successful outcome of the choice of strategy A in a risky transaction. Let $\psi^0 \in [0,1]$ be the actual probability of a successful outcome when choosing strategy A.

We assume that when an Investor analyzes a decision, his level of confidence is based on memories of the outcomes of his past projects.

An investor, in whose memory there are more successful investments, will have a higher level of confidence in the successful outcome of this situation. At the same time, the Investor, in whose memory there are more unsuccessful investments, will have a lower level of confidence in the successful outcome.

Let $s^0$ be the number of successful outcomes of strategy A in the memory of the Investor. And $f^0$ is the number of negative outcomes of strategy A.

We assume that the Investor confidence level is $p^0 = \beta(s^0, f^0)$, where β satisfies the following properties:

1) For any $s^0, f^0$ $0 < \beta(s^0, f^0) < 1$

2) $\beta'_s > 0$, $\beta'_f < 0$ (with an increase in the number of successes, confidence increases, and with an increase in the number of failures, confidence decreases)

3) For any $a > 0$, $\lim_{f \to \infty} \beta(af^0, f^0) = \frac{a}{a+1}$

The third property is fundamental, since it can be concluded from the fact that when the number of outcomes tends to infinity when choosing strategy A, the level of confidence tends to the empirical frequency of success.

Thus, the Investor forms in the memory the number of successful and unsuccessful outcomes. But we cannot say that he does it objectively, since it was previously stated that failures are perceived more deeply and are remembered more strongly. Investor's perception is distorted, which will necessarily affect our model. Consequently, we must formalize the description of the model and consider various options for the perception of the Investor.

## Model formalization. Variants of the distorted perception of the Investor.

Let $y^0 \in Y^0$ be the impulse received by the Investor, after each end of the project in which he invested. The investor remembers this impulse and records it in his past experience, which he will consider when choosing the strategy A. Also, let's say about the set of all possible outcomes of the transaction $X^0$.

We introduce the investor perception matrix. For each $y^0 \in Y^0$ and $x^0 \in X^0$ we denote $u^0_{x^0,y^0}$ - the probability that the momentum was $y^0$, when the true outcome is $x^0$. Since there are only two outcomes of the transaction, $X^0 = \{S^0, F^0\}$. And also only two types of impulse $Y^0 = \{n, m\}$, where n is the normal (correct) perception of the Investor, m is the false one. Let $\gamma^0$ be the probability that the Investor assigns a negative outcome to a negative state of affairs. Then we can create an investor perception matrix:

$$u^0 = \begin{pmatrix} & n & m \\ S^0 & 1 & 0 \\ F^0 & 1-\gamma^0 & \gamma^0 \end{pmatrix}$$

This matrix is important in order to assess the past experience of the Investor.

Let the Investor can accurately count the number of successful investments made (sometimes the memory may not allow him to do this), for which $x^0 = s^0$. Next, we formalize the concept of "experience" as a pair $w^0 = (x^0, y^0)$, as for each outcome $x^0$ there is a momentum $y^0$ (the investor responds to each outcome in a certain way, so a positive outcome is not always written by an investor in a positive experience). Due to the fact that a person does not always assess the situation objectively, at this moment there is a distortion of the actual data, and, therefore, we must introduce a random variable such as $c^0(w^0, s^0) = \{1, 0\}$. 1 if experience $w^0 \in s^0$ and 0 otherwise. Thus, $c^0 = 1$ if the experience that the Investor received (in his opinion) is related to a positive outcome.

Let the Investor recollection process be the vector $g^0 = \{g^0_{\omega^0, s^0}\}_{\omega^0, s^0}$, where $g^0_{\omega^0, s^0} = P\{c^0(w^0, s^0) = 1\}$ (the probability that the random variable $c^0 = 1$, as the experience obtained by the Investor is related to a positive outcome). Let also the random variable $\delta^0(s^0) = \sum_{w^0 \in W} c^0(w^0, s^0)$, where W is the set of all Investor's experiments.

Since the number of positive and negative outcomes deposited in memory for a given set of past experiments is a random variable, then $s^0 = \delta^0(S^0)$, $f^0 = \delta^0(F^0)$.

We can conclude that when the Investor forms the confidence level $p^0$, only those outcomes that he attributed to the normal (correct) one, ie those for which $y^0 = n$.

From where, the process of remembering an investor $g^0$ has the form:

$g^0_{\omega^0, s^0} = 1$ if $x^0 = s^0$ and $y^0 = n$

$g^0_{\omega^0, s^0} = 0$ otherwise.

Consider the frequency of positive implementations of strategy A, due to normal circumstances, according to Investor. This will be a function of the true frequency of outcomes when choosing strategy A.

Let $\alpha^0 = (\alpha^0_{x^0})_{x^0 \in X^0}$ be the true frequency of Investor outcomes.

The expected number of events when counting positive outcomes $s^0$ is

$$E_{(u^0, g^0, \alpha^0)} \delta^0(s^0) = \sum_{x^0 \in X^0, y^0 \in Y^0} g^0(x^0, y^0), s^0 \ u^0_{x^0, y^0} \ \alpha^0_{x^0}$$

Thus, the Investor's confidence level in investing $p^0$, based on past experience, is expressed by the following relationship:

$$\frac{E_{(w^0, g^0, \alpha^0)} \delta^0(s^0)}{E_{(w^0, g^0, \alpha^0)} \delta^0(s^0) + E_{(w^0, g^0, \alpha^0)} \delta^0(f^0)} \quad (1)$$

**Note:**

Formula (1) is an approximate calculation of Investor's confidence, since at any time t the number of successes in a transaction $s^0$ is a random variable. With an increase in past experience, its values will be close to the values of the relation (1).

**The investor's correct adoption of the strategy A:**

In order for the investor to invest his money in the project, it is necessary that

$$p^0 \geq \mu, \quad where\ \mu \in (0.5; 1]$$

Now we will consider such a scenario that the Investor can invest in the project not all the funds, but part of them. So, for example,

$$\mu^1 \in (0.5; 0.65]$$

$$\mu^2 \in (0.65; 0.8]$$

$$\mu^3 \in (0.8; 1]$$

Then if $\mu^1 \leq p^0 < \mu^2$, then the Investor invests 1/3 of the sum at his disposal, if $\mu^2 \leq p^0 < \mu^3$, then 2/3, and if $p^0 \geq \mu^3$, then the whole amount is full.

## Choosing a strategy for a project manager. Formation of the confidence level of the Project Manager

Now suppose that the Investor invests money in a project, and then the Investment Project Manager manages them. He will also have his strategy B: The manager can use the money for personal enrichment, instead of investing in the project. If the Manager uses money for personal enrichment, then he puts himself at risk of being caught by the police. Again, we ask ourselves the question: what is necessary for the Manager to consider in order to make a decision? Obviously, we must build on confidence and past experience, as is the case with Investor.

Let $p^1$ be the Manager's confidence level in the successful outcome of choosing strategy B in case of a risky decision to assign money such that $p^1 \in [0,1]$. Let $\psi^{01}$ be the actual probability of a successful outcome when the Manager chooses the strategy B, such that $\psi^{01} : [0,1] \times [0,1] \to [0, 1]$,

$$\psi^{01} = \psi(p^0, p^1) \in [0,1], \quad p^0 \in [0,1], \quad p^1 \in [0,1].$$

Let $s^1$ be the number of successful outcomes of the choice of strategy B in the Manager's memory, let $f^1$ be the number of negative outcomes of the choice of strategy B in the Manager's memory. We assume that the level of confidence of the Manager

$p^1 = \beta(s^1, f^1)$ where β satisfies the following properties:

1) for any $s^1$, $f^1$, $0 < \beta(s^1, f^1) < 1$

2) $\beta'_1 > 0$, $\beta'_2 < 0$

3) for any $a > 0$, $\lim_{f \to \infty} \beta(a f^1, f^1) = \frac{a}{a+1}$.

These assumptions can be viewed as the formalization of **accessibility** heuristics for the Manager.

## Formalization of accessibility heuristics due to distortion of the Manager's perception

Let $y^1 \in Y^1$ be the Manager's impulse, which he receives after each decision to assign funds. That is, choosing a strategy B, each time he receives a signal that corresponds to the perception of his current experience and which will affect the perception of the outcome of the choice of this strategy in the Manager's memory in the future. Let $X^1$ be the entire set of outcomes of the choice of the strategy B of the Manager.

We introduce the Manager's perception matrix. For each $y^1 \in Y^1$ and $x^1 \in X^1$ we denote $u^1_{x^1, y^1}$ - the probability that the momentum was $y^1$, when the true outcome is $x^1$. Since there are only two outcomes of a transaction, $X^1 = \{S^1, F^1\}$. And also only two types of impulse $Y^1 = \{n, m\}$, where n is the normal (correct) perception of the Manager, m is the false one. Let $\gamma^1$ be the probability that the Manager assigns a negative outcome to a negative state of affairs. Then we can create the Manager's perception matrix:

$$u^1 = \begin{pmatrix} & n & m \\ S^1 & 1 & 0 \\ F^1 & 1 - \gamma^1 & \gamma^1 \end{pmatrix}$$

For experience, we take the pair $w^1 = (x^1, y^1)$. Moreover, let $1_{w^1, s^1}$ be a random variable that takes a value of 1 if experience $w^1 = (x^1, y^1)$ included by the Superintendent in his calculation of past outcomes $s^1$ and equal to 0 otherwise. Since the Manager is not always able to remember all cases of previous outcomes.

Define a random variable $\delta^1(s^1) = \sum_{w^1 \in W^1} 1_{\omega^1, s^1}$

**The process of remembering** the Manager will be called the probability vector

$$q^1 = \{q^1_{w^1, s^1}\}_{w^1, s^1}$$

Where $q^1_{w^1, s^1} = P\{c^1(w^1, s^1) = 1\}$ (the probability that the random variable $c^1 = 1$, that is, the experience gained by the Manager, is related to a positive outcome).

Let $\alpha^1 = (\alpha^1_{x^1})_{x^1 \in X^1}$ be the true frequency of the Manager's outcomes when choosing the strategy B.

The expected number of events when counting the number of cases of successful outcomes $s^1$ is

$$E_{(u^1, q^1, \alpha^1)} \delta^1(s^1) = \sum_{x^1 \in X^1, y^1 \in Y^1} q^1_{(x^1, y^1), s^1} u^1_{x^1, y^1} \alpha^1_{x^1}$$

Thus, the Manager's confidence level for the personal use of $p^1$ based on past experience, is expressed by the following relationship:

$$\frac{E_{(w^1, q^1, \alpha^1)} \delta^1(s^1)}{E_{(w^1, q^1, \alpha^1)} \delta^1(s^1) + E_{(w^1, q^1, \alpha^1)} \delta^1(f^1)} \qquad (2)$$

**Note:**

Formula (2) is an approximate calculation of the Manager's confidence, since at any time t the number of successes in using the means $s^1$ is a random variable. With an increase in past experience, its values will be close to the values of the relation (2).

## Rule of adoption by the Manager

Let the use of funds for the project by the Manager for personal purposes entails possible costs, the level of which we denote as h. Suppose that the level of costs h is a stochastic and independent value from the interval [0,1]. And when the Manager considers options for strategy B, he knows the value of h.

If $p^1 \geq h$, that is, the Manager's confidence level in the successful outcome is greater than the required costs, then he takes strategy B and takes the money.

## 4. Conclusion

This work is an attempt to bring agents to the effective outcome of their interaction. Throughout the work, we were of the opinion (in common with psychology) that agents form their level of confidence based on simple heuristics and the choice of attribution style.

Undoubtedly, there are a large number of heuristics, psychological patterns and personal associations that agents use when forming their choice. But our main goal is not to compile an exhaustive catalog with which agents can form their determination, but to present a new way of optimization for the risk interaction between the two agents.

The results obtained in this article show that personal perception of the situation, a tendency toward optimistic or pessimistic distortions of perception can affect the formation of a person's confidence about his ability to succeed in a risky transaction. As well as the same mechanisms affect the opponent and how they are combined with each other.

From the obtained results a number of conclusions follow.

For example:

* if agents could find any control over their level of confidence, then this could lead them to improve the results of the risky deals they concluded.

* if agents could find any control over the level of confidence of the opponent, then this could lead them to improve the results of risky deals they make.

* Risky deals are "safer" to conclude with agents whose past experience is most positive.

Thus, careful attention to oneself and one's own reactions to the perception of situations, to the personality of the opponent and his past experience, is a very important step in the intention to conclude a risky deal.

**Acknowledgements**

The work is partly supported by work RFBR No. 18-01-00796.